\documentclass[12pt]{iopart}
\usepackage{dcolumn}   

\usepackage{graphicx,graphics,epsfig}   
\usepackage{dcolumn}    
\usepackage{bm}         
\usepackage{verbatim}   
\usepackage{color}      
\usepackage{times}
\usepackage{amsfonts,amssymb,graphics,graphics,color,times}

\DeclareMathAlphabet{\mathrsfs}{U}{rsfs}{m}{n}
\DeclareMathAlphabet{\mathpzc}{OT1}{pzc}{m}{it}
\DeclareMathAlphabet{\matheus}{U}{eus}{m}{n}
\DeclareMathAlphabet{\mathbbold}{U}{bbold}{m}{n}

\newcommand{\ba}{\begin{eqnarray}}
\newcommand{\be}{\begin{equation}}
\newcommand{\ee}{\end{equation}}

\newcommand{\ea}{\end{eqnarray}}
\newcommand{\ban}{\begin{eqnarray*}}
\newcommand{\ean}{\end{eqnarray*}}

\newcommand{\geneva}{\address{$^1$ D\'epartement de Physique Th\'eorique, Universit\'e de Gen\`eve, 1211 Gen\`eve, Switzerland}}
\newcommand{\paris}{\address{$^2$ LTCI, T\'el\'ecom ParisTech, 23 avenue dÕItalie, 75214 Paris CEDEX 13, France}}
\newcommand{\inria}{\address{$^3$ Inria, EPI SECRET, B.P. 105, 78153 Le Chesnay Cedex, France}}
\newcommand{\icfo}{\address{$^4$ ICFO-Institut de Ciencies Fotoniques, Mediterranean Technology Park, 08860 Castelldefels, Barcelona, Spain}}

\begin{document}

\title{Dimension of physical systems, information processing, and thermodynamics}

\author{Nicolas Brunner $^{1}$, Marc Kaplan $^2$, Anthony Leverrier $^3$, Paul Skrzypczyk $^4$}
\geneva
\paris
\inria
\icfo

\begin{abstract}
We ask how quantum theory compares to more general physical theories from the point of view of dimension. To do so, we first give two model independent definition of the dimension of physical systems, based on measurements and on the capacity of storing information. While both definitions are equivalent in classical and quantum mechanics, they are in general different in generalized probabilistic theories. 
We discuss in detail the case of a theory known as 'boxworld', and show that such a theory features systems with a dimension mismatch. This dimension mismatch can be made arbitrarily large by using an amplification procedure. Furthermore, we show that the dimension mismatch of boxworld has strong consequences on its power for performing information-theoretic tasks, leading to the collapse of communication complexity and to the violation of information causality. Finally, we discuss the consequences of a dimension mismatch from the perspective of thermodynamics, and ask whether this effect could break Landauer's erasure principle and thus the second law.
\end{abstract}

\maketitle

\section{Introduction}

Any theory aimed at explaining and predicting experimental observations makes use of a concept of dimension, which represents the number of degrees of freedom considered in the model. Loosely speaking, the dimension of a system corresponds to the number of perfectly distinguishable configurations the system can be prepared in. For quantum systems, this corresponds to the Hilbert space dimension.

The dimension of a physical system (classical or quantum) also characterizes the amount of classical information that can be encoded in the system and subsequently retrieved. Notably, classical and quantum systems of the same dimension $d$ can carry the same amount of classical information, namely $\log_2{d}$ bits of information \cite{holevo}. However, it is not the case that a classical system of dimension $d$ can always be substituted for a quantum system of the same dimension $d$. In fact, reproducing the behaviour of the simplest quantum system, the qubit, requires classical systems of infinite dimension \cite{galvao,hardy2}. This shows that quantum theory is much more economical in terms of dimension compared to classical physics \cite{gallego}. 
This is because classical and quantum systems are fundamentally different objects. On the one hand, for a classical system, the dimension $d$ always corresponds to the number of pure states the system can be prepared in. For finite dimension $d$, there are $d$ pure states. On the other hand, a quantum system, say a qubit, can be prepared in infinitely many different pure states, and two real numbers (hence infinitely many classical bits) are necessary to characterize a pure state. Indeed this difference has a strong impact on the capabilities of each theory for information processing.

The above shows that comparing classical and quantum physics from the point of view of dimension gives an interesting perspective on quantum mechanics, first as a physical theory but also on its power for information processing. 
More generally, one may ask how quantum mechanics compares to other physical theories, such as generalized probabilistic theories (GPTs) \cite{barrett}, a general class of theories featuring quantum and classical mechanics as special cases. In fact, understanding what features identify quantum mechanics among GPTs is a deep question, which may lead to a reformulation of quantum theory based on more physical axioms, see e.g. \cite{hardy,barrett,dakic,MM,chiribella,masanes,third}. Discussing information processing and physical properties of GPTs may give insight to this question \cite{barrett,janotta,steeg,murao,corsin,lal,serge}.

However, comparing different physical theories from the point of view of dimension is in general not a trivial issue. This is due to the fact that each theory has its own notion of dimension, usually based on the structure of the theory itself---e.g. in quantum mechanics, dimension is associated to the Hilbert space, the basic structure supporting quantum states and measurements.
But one may ask if there exists a universal, model-independent definition of dimension that could be used to compare different models. 

Here, we explore this problem, with the goal of finding out what is special about quantum mechanics as far as dimension is concerned. We start by discussing two definitions of dimension which we believe are natural. First we consider a notion of dimension related to the measurements that can be performed on a system. Our second notion of dimension refers to the amount of information that is potentially extractable from the system. While these two notions of dimension happen to coincide in classical and quantum physics, they are nevertheless different in general. 
We show that this is the case in a specific GPT known as 'boxworld' \cite{barrett}. We say that such a model features a \emph{dimension mismatch}. 

Moreover, the dimension mismatch of boxworld appears to be directly related to its astonishing information-theoretic power. So far the latter has been discussed in the context of spatially separated systems, that is considering communication tasks assisted by nonlocal (but no-signaling) correlations, see e.g. \cite{vanDam,NP,brassard,sandu,NLC,BS,IC,allcock,closed,ML,barnum,acin,GYNI,dani,LO}. In particular, it was shown that the existence of maximally nonlocal correlations (so-called Popescu-Rohrlich (PR) boxes \cite{PR}) would lead to the collapse communication complexity \cite{vanDam}, and to the violation of the principle of information causality \cite{IC}. Here we recover these results but following a different approach. Considering only single systems, we show that both the collapse of communication complexity and the violation of information causality, are direct consequences of the dimension mismatch of boxworld.

Finally we discuss the consequences of a dimension mismatch from a more physical perspective, related to thermodynamics. Specifically, we explore these ideas in the context of Maxwell's demon and Landauer's erasure principle \cite{landauer,bennett,koji}, and raise the question of whether theories with a dimension mismatch would break the second law of thermodynamics. Recent works \cite{koji2,barrettETH,esther} have also discussed thermodynamics in GPTs, however following a different approach.

\section{Two definitions of dimension}

We start by giving two model-independent definitions of dimension. While other definitions can indeed be considered, we believe the present ones are natural choices. 

\emph{Measurement dimension}, which we denote here $d_m$, is defined as the number of orthogonal pure states of the system, i.e. that can be perfectly distinguished in a single measurement \footnote{Although this may not hold in full generality, a possible alternative understanding of this definition is that $d_m$ is the number of outcomes of an ideal non-degenerate measurement. By an ideal measurement, or pure measurement, we refer to a measurement that is perfectly repeatable, i.e. when obtaining one particular outcome, repeating the measurement will always lead to the same outcome. By non-degenerate, we mean that the measurement is not coarse grained. In quantum mechanics, ideal non-degenerate measurements are rank-one projective measurements.}. Hence there is a set of states $\{\omega_1, \ldots, \omega_{d_m}\}$ and a measurement with $d_m$ outcomes, such that outcome $i$ has probability one for state $ \omega_i$. This implies that $\log_2(d_m)$ bits of information can be encoded in the system and subsequently retrieved via a measurement. This is definition of dimension represents arguably the most natural choice, and has been used previously, e.g. in $\cite{hardy}$.

The second definition, \emph{information dimension}, which we denote $d_i$, is the number of states of the system that are perfectly distinguishable pairwise. More formally, $d_i$ is the largest integer such that there exists a set of states $\{ \omega_1,..., \omega_{d_i}\}$ such that for any pair $ \omega_i$, $ \omega_j$ with $1\leq i<j \leq d_i$, 
there is a measurement that perfectly distinguishes $ \omega_i$ from $ \omega_j$. Note that these measurements can be different for all pairs of states, and can be chosen to be binary without loss of generality. In other words, information dimension characterizes the maximal number of states which are pairwise orthogonal. 

Finally, note that the measurement dimension is always smaller than or equal to the information dimension. Indeed, if a system can be prepared in $d_m$ different states which can be perfectly distinguished in a single measurement, then each pair of states in this set is also perfectly distinguishable by this measurement, hence $d_m \leq d_i$. Below we shall see that in quantum mechanics equality holds, i.e. $d_m = d_i$, while there exist more general theories where this is not the case, that is where $d_m < d_i$.

\section{Measurement and information dimensions coincide in classical and quantum mechanics}

We start with the case of quantum theory. Consider a quantum system, the states of which are given by vectors in a Hilbert space of dimension~$d$, i.e. $\mathbb{C}^d$. Then, any orthonormal basis consists of $d$ pure states. One can always define an observable acting on $\mathbb{C}^d$, formed by the $d$ states of an orthonormal basis. Thus, we get that $d_m=d$, that is the measurement dimension is equal to the Hilbert space dimension. 

It is also straightforward to see that the information dimension is equal to the Hilbert space dimension. Consider a set of $d$ quantum states which are perfectly distinguishable pairwise. These states are pairwise orthogonal and thus form an orthonormal basis in $\mathbb{C}^d$, hence we get that $d_i=d$. 

We conclude that for quantum systems, measurement dimension and information dimension coincide. Indeed, this is also the case for classical systems. A classical system of dimension $d$ features $d$ pure states, which are all pairwise distinguishable. On the other hand, a collection of classical states which are all pairwise distinguishable, can all be distinguished by a single measurement.

\begin{figure}
  \includegraphics[width=\columnwidth]{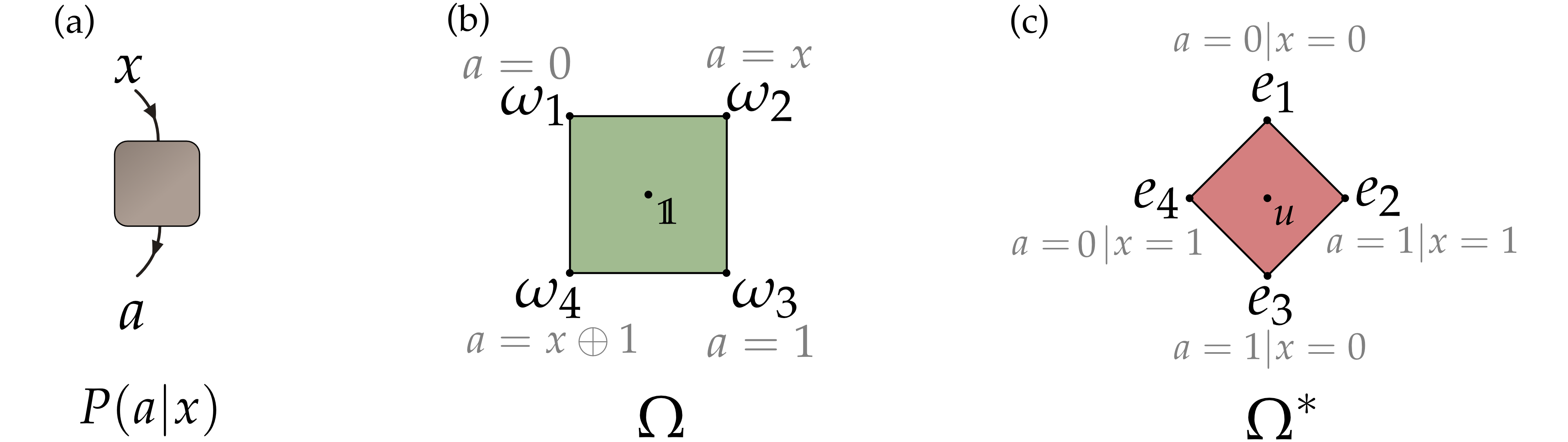}
  \caption{The simplest system in boxworld: the $g$-bit. (a) This system can be understood as a black box taking a binary input $x=0,1$ and returning a binary output $a=0,1$. The state of the system is described by a conditional probability distribution $P(a|x)$. (b) The state space $\Omega$ of the system can be represented as a square in $\mathbb{R}^2$. The system thus features four pure states, $\omega_j$. For each pure state, the outcome $a$ is a deterministic function of the input $x$, as indicated. At the centre of $\Omega$ is the maximally mixed state, that is, where $a$ is independent of $x$ and random. (c) The space of effects, $\Omega^*$, is the dual of $\Omega$. It features four extremal effects, $e_j$, which correspond to the four measurement outcomes, i.e. obtaining output $a$ for a given input $x$. The probability of $e_j$ on any state is easily determined. For instance, effect $e_1$ has probability one for states $\omega_{1,2}$ and probability zero for states $\omega_{3,4}$. There are two pure measurements for this system: the first is composed of extremal effects $e_1$ and $e_3$, hence corresponding to input $x=0$; the second is composed of extremal effects $e_2$ and $e_4$, hence corresponding to input $x=1$. Note that $e_1+e_3=e_2+e_4=u$, where $u$ is the unit effect.}
\label{fig1}
\end{figure}

\section{Dimension mismatch in generalized probabilistic theories}

While it may come to no surprise that measurement dimension and information dimension coincide in quantum mechanics, we shall see that this is not necessarily the case in generalized probabilistic theories. 
When measurement dimension and information dimension differ, i.e. when $d_m < d_i$, we say that the theory features a \emph{dimension mismatch}. Before discussing this aspect, we give a brief review of the concepts of GPTs; more comprehensive reviews are available in the literature, see e.g. \cite{barrett,corsin,janotta,wilce}.

\subsection{Basic concepts}

The state of a system is an abstract object that defines the outcome probabilities for all the measurements that can be
performed on a system. The state space $\Omega$ is the set of states that a system can be prepared in. The natural requirements that the model is no-signaling (i.e. that it does not allow for instantaneous transmission of information), and that probabilistic mixtures of states are also valid states, enforce convexity and linearity of $\Omega$. $\Omega$ (assumed here to be closed) thus lives in a vector space. A state corresponds to a vector $\omega \in \Omega$. Pure states are extremal points of $\Omega$. All other states are called mixed states, as they can be decomposed as a convex mixture of pure states. 

Consider now measurement outcomes, usually termed \emph{effects}. These are affine maps $e(\omega)$ from $\Omega$ to the interval $[0,1]$, hence assigning a probability to each state $\omega$. A measurement is then given by a set of effects $\{e_i\}$ such that $\sum_i e_i = u$, where $u$ is called the unit effect, i.e. the effect assigning value 1 to every state in $\Omega$.
This condition ensures that all states are normalized. 
Similarly to quantum theory, a measurement is given by a decomposition of the unit effect (the identity), ensuring that the probabilities for each outcome sum to one. The set of valid effects is given by $\Omega^* =  \{ e : 0 \leq e(\omega) \leq 1 \,\,\forall \omega \in \Omega \} $, which is the convex hull of extremal effects, the unit effect, and the zero effect.

Let us now discuss our above notions of dimension in the context of GPTs. 
We focus here on a specific GPT known as boxworld \cite{barrett}, studied later in e.g. \cite{barrettshort,colbeck,short2013}. An interesting aspect of this model is that it features nonlocal correlations (i.e. leading to violation of Bell inequalities); in fact, boxworld features all possible no-signaling correlations, such as the PR box. 
We start by discussing the simplest system in boxworld. 
This is a single physical system on which two possible binary measurements can be performed. Such a system is usually represented as a 
``black box" with two possible inputs (or pure measurements), denoted $x=0,1$, and two possible outputs (or measurement outcomes), denoted $a=0,1$ (see Fig.1a). The pure states of this system are those for which the outcome $a$ is a deterministic function of the input $x$, that is 
\ba a = \alpha x \oplus \beta \ea
where $\alpha,\beta = 0,1$ and $\oplus$ denotes addition modulo 2. There are thus four pure states. Using the vectorial representation of GPTs, the state space $\Omega$ of this simple system can be conveniently represented by a 2D cube, i.e. a square (see Fig.1b). The vertices of the square represents the four pure states $\omega_j$ with $j=1,...,4$. All other states are convex mixtures of the pure states, and are thus mixed states. The set of effects $\Omega^*$, given by the dual of the state space, is also a 2D cube, but rotated compared to the state space (see Fig.1c). Pure (or extremal) effects correspond to the four possible measurement outcomes, i.e. inputting $x$ in the box and getting outcome $a$. For more details on this system, in particular on its state space, we refer the reader to Refs \cite{barrett,janotta}.

Clearly, our system features only two ideal measurements, which correspond to input $x=0$ or $x=1$ in the box. Both measurements have two outcomes, hence we have that measurement dimension is $d_m=2$. 
This system can be used to transmit exactly one bit of information, and thus can be considered as a generalization of the qubit---in fact it is often referred to as a $g$-bit, i.e. generalized bit~\cite{barrett}. Moreover, such a system can be considered as the reduced local part of a PR box.

On the other hand, we see immediately that the information dimension is $d_i=4$, since any pair of pure states can be perfectly distinguished by performing either the $x=0$ or $x=1$ measurement. Hence we see that our $g$-bit system features a dimension mismatch, as $d_i>d_m$.

\subsection{Amplification of dimension mismatch}

While the $g$-bit has a dimension mismatch of a factor 2, we will see now that there exist in boxworld systems with an arbitrarily large dimension mismatch. Such systems, which are a straightforward generalization of the $g$-bit, can in fact be constructed by composing $g$-bits. Hence, dimension mismatch can be amplified.

We presented the $g$-bit system as a black box featuring two possible (pure) binary measurements $x=0,1$, with outcome $a=0,1$. Consider now a black box with $D$ possible binary measurements $x=0,\ldots,D-1$, with outcome $a=0,1$. Following the $g$-bit construction, the state space of our system is now a hypercube in $\mathbb{R}^D$. The system features $2^D$ pure states. Each state associates a deterministic outcome ($a=0$ or $a=1$) to each of the $D$ possible measurements. In the state space, each pure state corresponds to one of the vertices of the $D$-dimensional hypercube, and can therefore be conveniently described by a vector of the form $\{\zeta_i\}_{i=1...D}$, where $\zeta_i=0,1$ denotes the outcome obtained when performing measurement $i=1...D$.

The measurement dimension is $d_m=2$ for all $D$, since all pure measurements are binary. Hence, the system can be viewed as a generalization of a bit, since only one bit of information can be retrieved by performing a measurement on it. From now on, we will thus refer to this system as a \emph{hypercube bit} \footnote{Note that the hypercube bits discussed here are different objects than the 'hyperbits' discussed in M. Pawlowski and A. Winter, Phys. Rev. A {\bf 85}, 022331 (2012).}. 

On the other hand, any pair of pure states of the hypercube bit can be perfectly distinguished, since there is (at least) one measurement $x$ for which two different pure states give different outcomes. Therefore, the information dimension is $d_i=2^D$. Hence, as $D$~increases, we obtain an arbitrarily large dimension mismatch.

Finally, we present a procedure to construct hypercube bits by composing $g$-bits. To define the composition of systems in a GPT, one has to choose a tensor product. In boxworld, one uses the 'maximal tensor product' \cite{barrett} (see also \cite{janotta}). Intuitively, this means the following: a bipartite (or more generally multipartite) system is considered as a valid state, if (i) it does not allow for instantaneous transmission of information (i.e. it is no-signaling) and (ii) upon performing a measurement on one system, and for each outcome of this measurement, the other system is prepared (or steered) into a state that is valid (i.e. inside the initial state space) \footnote{More formally, the set of bipartite states in the maximal tensor product can also be constructed by first defining the set of bipartite effects. The extremal bipartite effects are taken to be of the product form $e_i \otimes e_j $, where $e_{i,j}$ are extremal effects for a single system. In other words, we combine effects using the minimal tensor product. Finally, we consider all bipartite states that can be consistently defined on all these bipartite effects.}. 

Consider the composition of two $g$-bits. The resulting system can be represented as a black box shared between two observers, such that each observer can perform one out of two possible measurements with binary outcomes. We denote $x_i$ the measurement of observer $i$ and $a_i$ its outcome. The state space obtained by taking the maximal tensor product of two $g$-bits has been extensively discussed \cite{barrett05}. For our purpose, it is convenient to express the pure states of this bipartite state space according to their correlations: 
\ba\label{corrs} a_1 \oplus a_2 = \alpha x_1 x_2 \oplus \beta x_1 \oplus \gamma x_2 \oplus \delta  \ea 
where $\alpha,\beta,\gamma,\delta = 0,1$. We thus obtain 16 pure states \footnote{Note that when including marginals, the state space features 24 pure states: (i) 16 local deterministic states, and (ii) 8 nonlocal PR boxes (see \cite{barrett05} for details). However, projecting onto the correlations subspace, i.e. considering only the correlations $a_1 \oplus a_2$, the 16 deterministic states are mapped into 8 states with deterministic correlations.}, which can be divided into two classes: (i) 8 local deterministic states, for which $\alpha=0$, and (ii) 8 nonlocal PR boxes, for which $\alpha=1$. Note that states in class (i) can be obtained by taking two copies of a $g$-bit: hence $a_i = f(x_i)$. On the contrary, states in class (ii) feature nonlocal correlations (e.g. of the form $a_1 \oplus a_2 = x_1 x_2$) which achieve maximal violation of the Clauser-Horne-Shimony-Holt Bell inequality. To ensure no-signaling, such states have full local randomness, that is $p(a_i=0|x_i)=1/2$.

At this point, our composed system has $d_m=4$. To obtain a hypercube bit, we project it onto a subspace. More precisely, we apply the projection $(a_1,a_2) \rightarrow a= a_1 \oplus a_2$. This is analogous to parity projections in quantum theory. The resulting system has measurement dimension $d_m=2$ and information dimension $d_i=16$, and is in fact isomorphic to a hypercube bit with $D=4$.

The above procedure can be applied starting from the composition of $k$ $g$-bits, hence resulting in a system that is isomorphic to a hypercube of dimension $D=2^k$. Specifically, one considers all pure states in the maximal tensor product of $k$ $g$-bits with correlations of the form $a_1 \oplus a_2 \oplus ... \oplus a_k = f(x_1,...,x_k)$ where $f$ is an arbitrary deterministic boolean function of $k$ bits---note that in general there exist additional pure states which are not of this form \cite{pironio}. Hence we obtain $2^{2^k}$ pure states. After performing the projection $(a_1,...,a_k) \rightarrow a= a_1 \oplus a_2 \oplus ... \oplus a_k$ we get a state space that is isomorphic to the hypercube of dimension $2^k$. Therefore, when composing $g$-bits using the maximal tensor product (in other words assuming only no-signaling), we obtain systems with an arbitrarily large dimension mismatch. 

Finally, note that the possibility of amplifying dimension mismatch depends on the way systems are composed, i.e. which tensor product is chosen. Since the choice of tensor product will affect the strength of nonlocal correlations featured in the model, there appears to be a direct relation between the possibility of amplifying dimension mismatch and the strength of nonlocal correlations. This aspect will be discussed in more detail in Section \ref{discussion}.

\section{Dimension mismatch and communication power}

In this section, we investigate the consequences for information processing of the dimension mismatch of hypercube bits. We consider information-theoretic tasks in GPTs using protocols with 'one-way communication', where Alice sends a system to Bob \cite{barrett}. Specifically, we consider the situation in which Alice can prepare a hypercube bit in any desired state, encoding certain information in it. Then she sends the system to Bob, via an adequate (non-classical) channel. Bob finally performs a measurement on the received hypercube bit, which allows him to extract part of the information that Alice encoded in the system. 

We shall see that the dimension mismatch of hypercube bits enhances their communication power compared to classical and quantum resources, leading to maximal violation of information causality and to the collapse of communication complexity. In fact, these enhancements in communication power are captured by the fact that hypercube bits are tailored for computing the index function, which we discuss first.

\subsection{Index function}

Consider the following communication complexity problem. Alice receives a uniformly sampled bit string of length $n$, $\mathbf{b}=b_1 \ldots b_n$. Bob receives a random index $k \in \{1, \ldots ,n \} $. The goal for Bob is to output the value of the bit $b_k$. This problem is known as computing the index function. Its classical and quantum communication complexity is $n$~bits \cite{nayak}. Hence, using classical or quantum systems for the index function requires systems of dimension increasing exponentially with $n$. 
Below we give a protocol which uses one hypercube bit of dimension $D=n$ and computes the index function on inputs of length $n$. Since we have here that $d_m=2$ independently of $n$, this is in stark contrast with classical and quantum resources. Note that a similar protocol was presented in \cite{barrett}, however using $N$ gbits (with $d_m=2^N$) instead of a single hypercube bit.

After receiving her input bit string $\mathbf{b}=b_1 \ldots b_n$, Alice prepares a hypercube system of dimension $D=n$ in a pure state $\{\zeta_i\}_{i=1...D}$ such that $\zeta_i=b_i$. Hence the outcome of measurement $x_i$ is $\zeta_i=b_i$ (for $i=1,...,D$) . Then she sends the system to Bob. Upon receiving index $k$, Bob performs measurement $x_k$ and obtains outcome $\zeta_k$. He thus retrieves correctly the bit value $b_k$.

Note that the ability to compute the index function depends only on the information dimension $d_i$, but is independent of the measurement dimension $d_m$. For hypercube bits, the index function can be computed for arbitrary $n$ (taking $D=n$). Nevertheless, the measurement dimension remains fixed, $d_m=2$, meaning that the system is binary, in the sense that only one bit of information can be retrieved by performing a measurement on it. This is thus in stark contrast with systems featuring no dimension mismatch, such as classical and quantum systems.

\subsection{Information causality}

The principle of information causality \cite{IC} was initially formulated in a scenario where classical communication is assisted with pre-shared non-local correlations. The principle limits the strength of nonlocal correlations by restricting the power of classical communication assisted by nonlocal correlations. In certain cases, information causality allows one to recover the boundary between quantum and super-quantum correlations \cite{IC,allcock}.

Here we consider a variant of information causality, in a scenario with one-way communication. Alice encodes information in a physical system (for instance a hypercube bit) and sends it to Bob, who retrieves information by performing a measurement on the system. In this context, the principle should limit Bob's conditional gain of information depending on the information capacity of the physical system that is sent. Consider Alice receiving a bit string of length $n$, $\mathbf{b}=b_1 \ldots b_n$, and Bob receiving an index $k \in \{1, \ldots ,n \} $ and asked to give a guess $\beta$ for the value of bit $b_k$. Then information causality can be written as 
\ba  \label{IC} \sum_{j=1}^n I(b_j:\beta |k=j) \leq H(d_m), \ea 
where $d_m$ is the measurement dimension of the system and $H(d_m)$ is its capacity, that is the amount of bits of information that be retrieved from the system by performing a measurement on it \footnote{Note that Czekaj and colleagues \cite{horo}Êrecently proposed a physical principle for single systems similar to our reformulation of information causality, see equation (3). Interestingly these authors found that various classes of GPTs led to a violation of this principle.}.

Indeed, this task is essentially identical to the index problem discussed above. Hence by using a hypercube bit of dimension $D=n$, Alice and Bob can achieve $\sum_{j=1}^n I(b_j:\beta |k=j) = n > 1= H(d_m)$ and violate maximally information causality. In other words, one can implement a perfect 1-out-of-$n$ random access code \cite{nayak} using a hypercube bit of dimension $D=n$. 

Note that a related formulation of information causality for single systems was recently considered in an axiomatic reformulation of quantum theory \cite{masanes}.

\subsection{Collapse of communication complexity}

The collapse of communication is arguably the strongest demonstration of the communication power of PR boxes \cite{NP,sandu}. Initially proved for pure PR boxes \cite{vanDam}, the result was later extended to classes of noisy PR boxes \cite{brassard,BS}. 
While these works considered classical communication assisted with pre-shared nonlocal resources, we consider here the case where Alice prepares a physical system and sends it to Bob. 

Consider a boolean function $f: X \times Y \rightarrow \{0,1\}$, where $X$ denotes the set of possible bit strings $\mathbf{b}=b_1 \ldots b_n$ received by Alice, and $Y$ the set of bit strings $\mathbf{c}=c_1 \ldots c_m$ for Bob. The goal is that Bob should output the value of the function $f(\mathbf{b},\mathbf{c})$, for any possible inputs. The communication complexity of the function is then the amount of information that needs to be communicated from Alice to Bob in order for Bob to compute the function. Classically, this represents the number of bits $C$ required by the most economical protocol, hence this protocol uses classical systems of dimension $d=2^C$. Using quantum resources, the communication complexity is the number of qubits $Q$ necessary to compute the function, hence the optimal protocol requires quantum systems of Hilbert space dimension $d=2^Q$. More generally, we see that a protocol using a system of measurement dimension $d_m$ has communication complexity $C=\log_2(d_m)$, as this represents the amount of information that is extracted from the system by Bob's measurement.

By adapting the index protocol discussed above, any boolean function can be computed by having Alice sending a single hypercube bit to Bob. The protocol is as follows. Suppose Alice and Bob want to compute a given function $f$. Upon receiving her input bit string $\mathbf{b}$, Alice computes locally the value of the function $f(b,c)$ for all possible input bit strings $\mathbf{c}$ of Bob. Next, she prepares a hypercube bit of dimension $D=\vert Y \vert=m$ (i.e. the number of possible inputs $y$ for Bob) in the state $\{\zeta_i\}_{i=1,...,D}$ such that $\zeta_i = f(\mathbf{b},\mathbf{c}=i)$. Hence, we need a hypercube bit of sufficiently large dimension, such that $D=m$. Bob, upon receiving his bit string $\mathbf{c}$, performs the corresponding measurement and accesses the bit $\zeta_c$, hence the value of the function $f(\mathbf{b},\mathbf{c})$.

Since the above protocol uses a single hypercube bit, it has trivial communication complexity $C=1$, independently of the size of the input bit strings. Indeed, this is in stark contrast with the case of classical or quantum resources, for which there exist functions whose communication complexity is not trivial, that is, $C$ is an increasing and unbounded function of the size of the problem \cite{cleve}.

\subsection{Steering}

We just showed that the existence of hypercube bits would lead to the violation of information causality and to the collapse communication complexity. Hence we recover previously known results about the dramatic enhancement of communication power observed in boxworld compared to quantum theory \cite{IC,vanDam}. However we followed a different approach, similar to considerations made in Ref. \cite{barrett}. While previous works \cite{IC,vanDam} considered the power of classical communication assisted by the strong nonlocal correlations (i.e. PR boxes) available in boxworld, we studied here the power of hypercube bits for information-theoretic tasks, considering protocols in which Alice prepares a hypercube bit and then sends it to Bob, who finally performs a measurement on it. It turns out however that both approaches can be connected, as the communication of a hypercube bit is essentially equivalent to the communication of one classical bit assisted with PR boxes, as we shall see below.

To see this, consider first the case in which Alice prepares a hypercube bit of dimension $D$ in a state $\{\zeta_i\}_{i=1,...,D}$ and sends it to Bob who recovers bit $b_k$ by performing measurement $k$. This protocol can be simulated by $D$ PR boxes, shared between Alice and Bob, and one bit of classical communication from Alice to Bob. We denote the inputs (outputs) of PR box number $i$ by $x_i$ ($a_i$) for Alice and $y_i$ ($b_i$) for Bob. First, Alice enters bit $\zeta_i$ in PR box number $i$ for $i=1,...,D$. She then sends to Bob the bit $c=a_1 \oplus ...\oplus a_D$. Bob, who wants to retrieve bit $k$, inputs $y_k=1$ and $y_j=0$ for $j \neq k$. Finally, Bob outputs $ c \oplus b_1 \oplus... \oplus b_D = \zeta_k$.

Consider now the converse problem. We want to simulate a situation where Alice and Bob share $n$ PR boxes, Alice sends one bit of classical communication to Bob who then extracts one bit of information $f(\mathbf{b},\mathbf{c})$ which depends on the inputs $\mathbf{b}$ and $\mathbf{c}$ they received. It is sufficient here to consider a deterministic function~$f$. Upon receiving her input $b$, Alice computes locally $f(\mathbf{b},\mathbf{c})$ for all possible inputs of Bob $c=1,...,m$. She then prepares a hypercube bit of dimension $D=m$ in a state $\{\zeta_i\}_{i=1,...,D}$, such that $\zeta_i= f(\mathbf{b},\mathbf{c}=i)$. Upon receiving the hypercube bit from Alice, Bob performs measurements to retrieve the bit $\zeta_c$, and retrieves the desired bit $f(\mathbf{b},\mathbf{c})$, as he would have in the situation they wanted to simulate.

Thus, we conclude that the communication of a hypercube bit can be replaced by a single bit of classical communication assisted by a sufficient number of PR boxes, and vice versa. 
The fact that the hypercube bit has measurement dimension $d_m=2$ corresponds to the fact that a single bit of classical communication is sufficient in the above protocol. Also, the information dimension $d_i$ is related to the number of required PR boxes in the protocol we gave above.

\section{Discussion}\label{discussion}

We have discussed the problem of characterizing the dimension of physical systems in a model independent way. We introduced two definitions for the  dimension which we believe are rather natural. One is related to the number of outcomes of a pure measurement and the other to the number of pairwise perfectly distinguishable states. These two notions of dimension coincide in quantum mechanics and are equal to the Hilbert space dimension, the usual (but model-dependent) measure of dimensionality of quantum systems. We showed that this is not the case in certain generalized models, such as boxworld. Hence these models feature a dimension mismatch, in the sense that the number of states which are pairwise perfectly distinguishable exceeds the number of outcomes of a pure non-degenerate measurement. Moreover, we described a procedure to amplify the dimension mismatch, leading to an arbitrary large mismatch. We then investigated the link between dimension mismatch and communication power, showing violation of information causality and the collapse of communication complexity in boxworld.

An interesting point is that we recover here previous results on the astonishing communication power in boxworld, but using a different approach. Instead of considering two observers sharing pre-established nonlocal correlations and sending classical communication to each other, we considered the case in which one observer sends a physical system (a hypercube bit) to the other. Hence it is the properties of a single system which are exploited here, instead of the strong nonlocal correlations of bipartite systems. This illustrates the strong connection that exists between the properties of the state space of a single system with the strength of nonlocal correlations that are obtained when looking at bipartite systems \cite{barnum,acin,janotta,gonzalo}. It is important to note that, although we do not directly consider bipartite (or multipartite) systems, we did make an assumption about how single systems can be combined. Specifically, in the procedure we used to amplify dimension mismatch, it was essential to assume that systems can be combined using the maximal tensor product, which does ensure the existence of maximally nonlocal correlations, such as PR boxes. This shows that dimension and nonlocality are connected.

Another aspect worth mentioning is that concepts of dimension discussed here allow us to characterize information processing in GPTs without introducing a notion of entropy, a notoriously problematic issue \cite{wehner,barnum2,dahlsten}.

It would be interesting to see how general the present results are. Considering an arbitrary model featuring a dimension mismatch, can this dimension mismatch always be amplified? What are the information-theoretic properties of such a model? For instance, does a dimension mismatch always lead to a collapse of communication complexity? Another possible direction is to consider the present ideas for the problem of deriving quantum theory from a set of physically reasonable axioms \cite{hardy,barrett,dakic,MM,masanes}. Indeed, imposing that measurement dimension and information dimension are equal seems to have nontrivial consequences on the structure of the underlying physical theory and its information-theoretic capabilities. In order to address these questions, it would be desirable to adapt the present definitions of dimension. Specifically, instead of asking for perfect distinguishability (as we do here), one could impose a lower bound on the distinguishability of states, choosing an appropriate measure of distinguishability. Note that a recent work \cite{horo} suggested to use the mutual information in this context, proposing a principle similar to our reformulation of information causality; see eq. (3). We expect such an approach to detect a considerably larger class of theories featuring a dimension mismatch. 

Another direction would be to see whether the present approach is related to the principle of local orthogonality \cite{LO,LO2}, also referred to as the exclusivity principle \cite{adan} when discussing contextually. The main idea behind this principle is to demand the following. Consider a set of events of obtaining outcomes $a_1 \ldots a_n$ upon performing measurements $x_1 \ldots x_n$, which we denote by $(a_1 ... a_n |x_1 ... x_n)$. Here we focus on sets of events where all possible pairs of events are orthogonal, i.e. mutually exclusive (at least one party performs the same measurement but obtains different outcomes). The principle then says that the sum of conditional probabilities of all events (in the set) should not exceed one. That is, events which are pairwise orthogonal are all mutually exclusive. Indeed, the notion of pairwise orthogonality that appears here is reminiscent of our definition of the information dimension, which represents the maximal number of states which are all pairwise orthogonal. We conjecture that, in a model without dimension mismatch, that is where $d_m=d_i=d$, the principle of local orthogonality should be satisfied. Consider a set of $d$ pairwise orthogonal events, i.e. measurement outcomes or effects. Then one should be able to construct from these effects a $d$-outcome measurement, which is pure and non-degenerate. Indeed, such a measurement must be properly normalized, i.e. the sum of the probabilities of each outcome must be one. Hence, if such a measurement can always be constructed, it follows that local orthogonality is satisfied. It would be interesting to derive a formal proof from the above intuitive argument. The converse link also deserves attention, that is, does local orthogonality imply no dimension mismatch, i.e. $d_m=d_i=d$? Note that an interesting step in this direction was recently given in Ref. \cite{LO3}, where it is shown that the set of 'almost quantum correlations' \cite{almostQ} (i.e. a superset of quantum correlations which is proven to satisfy local orthogonality and that does not allow for trivial communication complexity) will have no dimension mismatch.

\section{Thermodynamics perspective}

Finally, we believe that the present ideas also have implications from a more physical perspective, in particular for thermodynamics. The main point we would like to raise here is that a dimension mismatch may potentially lead to a violation of the second law. We formalize this question by asking if a system with a dimension mismatch, such as the hypercube bit, could lead to the realization of Maxwell's demon. Note that recent works have also raised discussed thermodynamics in the context of GPTs \cite{koji2,barrettETH,esther}. 

We first argue that the concept of measurement dimension is related to the cost of erasing the information encoded in a physical system. For instance, in quantum theory, the cost of erasing a qudit (a system in a Hilbert space of dimension $d$ and a vanishing Hamiltonian) is given by $k_B T \log(d)$ (see e.g. \cite{koji}), where $T$ is the temperature of the external bath used to perform the erasure, and $k_B$ is the Boltzman constant. 
This can be done by performing an ideal and non-degenerate measurement on the system and then erasing the information about the outcome of this measurement, which costs exactly $k_B T\log(d)$. Note that the state of the system after the measurement is one of the $d$ pure eigenstates of the measurement. Hence, one can finally reset the system back to any desired pure state by applying a suitable unitary transformation to the system---this operation being unitary does not change the entropy of the system, and since the Hamiltonian vanishes it costs no energy.

\begin{figure}
  \includegraphics[width=0.7\columnwidth]{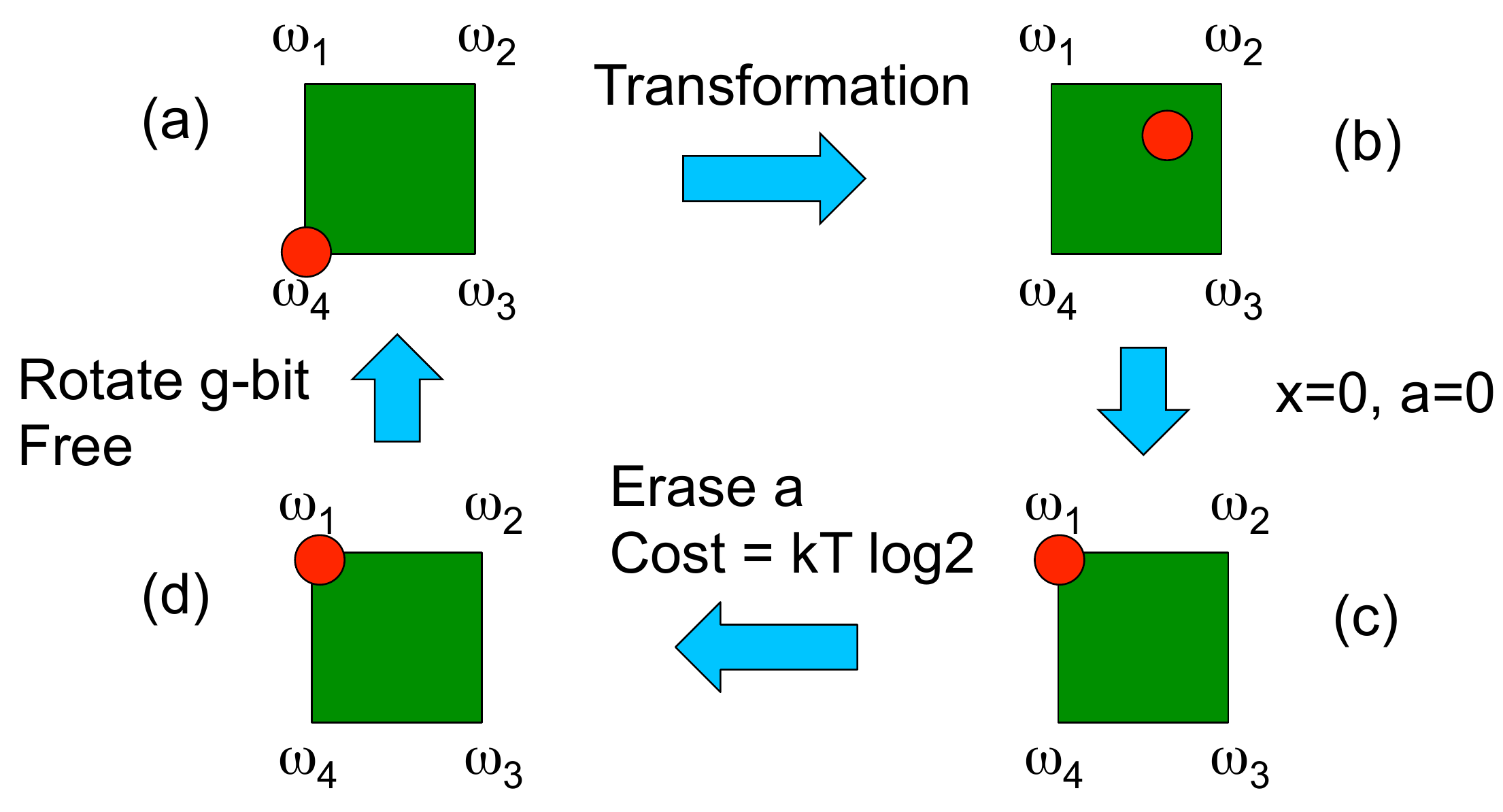}
  \caption{Cost of erasing a $g$-bit. We present a simple cycle for erasing the information contained in a $g$-bit. (a) Consider a $g$-bit in some fixed initial state $\omega_4$ (as indicated by the red dot). (b) After an arbitrary transformation, during some (unknown) information may have been encoded in the $g$-bit, the $g$-bit is an arbitrary and unknown state. (c) In order to erase the information contained in the $g$-bit, one starts by performing a measurement, here $x=0$, giving an outcome, say $a=0$. According to the measurement dynamics described in the main text, i.e. the procedure we adopted for defining the post-measurement state, the $g$-bit is now in the pure state $\omega_1$. (d) To bring the system in the initial state, one must erase the (classical) information about the measurement outcome $a$, now stored in a classical register. Having access to a thermal bath at temperature $T$, the energy cost of this operation is $k_BT \log{2}$. Finally, the state of the $g$-bit must be reset to the pure state $\omega_4$. Since all pure states have here the same energy, this operation can be done for free. Hence the cycle is complete. Finally, notice that the above protocol can be straightforwardly generalized to erase a hypercube-bit. The energy cost of such a procedure is $k_BT \log{d_m}= k_B T \log{2}$.}
\label{fig2}
\end{figure}

Following this line of reasoning, we will now argue that the cost of erasing a hypercube bit is $k_BT \log d_m = k_BT \log 2$. 
An aspect that deserves clarification here is how one should define the post-measurement state in the case of a hypercube bit. For simplicity, let us focus on the g-bit. We assume here that the post-measurement state for measurement $x=0$ is given by the pure state $\omega_1$ when $a=0$, and $\omega_3$ when $a=1$. Similarly for measurement $x=1$, we assume that the post-measurement state is $\omega_2$ when $a=0$, and $\omega_4$ when $a=1$. Indeed, this choice is somehow arbitrary, and it would be possible to define the post-measurement states differently. Nevertheless, our choice is a valid one, and is consistent with the model, to the best of our knowledge. Note also that when performing one measurement, say $x=0$, the information about the other measurement outcome, $x=1$, is  erased, ensuring that only one bit of information can be retrieved by performing a measurement on the $g$-bit.

Now let us investigate the cost of erasing a $g$-bit. Similarly to the quantum case, one first performs a measurement, say $x=0$. Then the $g$-bit will be found in state $\omega_1$ or $\omega_4$ (depending on the outcome of the measurement) and can then be reset to any desired pure state by applying a suitable rotation of the square (equivalent to a unitary operation). Finally, the information about the measurement outcome must be erased, which costs $k_BT \log(2)$. Generalizing this procedure to hypercube bits, one thus gets that the cost of erasure of a hypercube bit is $k_BT \log d_m = k_B T \log 2$.

Next, one may wonder what is the significance of the information dimension in this context. It is tempting to conjecture that the information dimension captures the amount of information that the system can store, i.e its memory size. Consider Maxwell's original thought experiment, in which a demon separates fast and slow particles, hence making heat flow from a cold bath into a hot one---for more details, see \cite{bennett,koji}. For each incoming particle, the demon, after determining the speed of the particle, decides to let it cross the partition or not. The demon must then store this information in his memory. 

Now let us imagine that the demon's memory is in fact a hypercube bit of dimension $D$. We denote its state $\{\zeta_i\}_{i=1,...,D}$. Say the hypercube bit is initially in the state $(\zeta_1=0,\zeta_2=0,...,\zeta_D=0)$. After deciding on the fate of the first incoming particle, the demon stores the information in the hypercube bit: if he decided to let the particle pass (0), the demon leaves the state of the hypercube unchanged; but if he decides to bounce the particle back (1), he rotates the hypercube bit to the state $(\zeta_1=1, \zeta_2=0,...,\zeta_D=0)$, that is he flips the first bit of the state. The demon then proceeds similarly for the next $D-1$ incoming particles, by updating the state of the hypercube such that the bit of information created in the $k$-th run is stored in $\zeta_k$. Notice that all pure states of the hypercube-bit are assumed here to have the same energy, hence updating the state does not cost work. Moreover, note that the demon should rotate the hypercube-bit in a different direction in each step of the protocol, and erase in the final step. He can keep track of the current step of the protocol by using a classical clock, and acting deterministically in each step.

It thus seems that the demon can store $D$ bits of information in the hypercube bit. After that, in order to continue sorting the particles, the demon must erase the hypercube bit, and make sure to reset it to the initial state $(\zeta_1=0,\zeta_2=0,...,\zeta_D=0)$. According to Landauer's erasure principle \cite{landauer}, this operation should cost him at least $k_BTD \log(2)$. However, we have seen above that the cost of erasing a hypercube bit (and resetting it to an arbitrary pure state) is in fact much smaller, namely $k_B T\log d_m = k_BT \log 2$. Hence we get a contradiction with Landauer's principle, thus implying a violation of the second law.

At this point it is legitimate however to question the significance of the information dimension here. In particular, one may argue that, since only one bit of information can be retrieved from the hypercube bit, the memory size of such a system is in fact 2, and not $D$ as argued above. However, one is forced to admit that all the information about which particles the demon let pass and which ones he did not let pass is encoded in the hypercube bit, and that any single bit of this information can be retrieved at any moment. Moreover, for each new bit of information to be stored, the system is either left in the same state (if the bit to be stored has value 0) or flipped to an orthogonal state (if the bit value to be stored is 1). Another aspect that one may question, is our choice of post-measurement states. In fact, one may argue that the above argument gives evidence that such a choice of post-measurement state is not a valid one. However that would imply postulating the validity of the second law (rather than deriving it), which is not desirable. An interesting question would be to revisit the above argument with other choices of post-measurement states, for instance choosing post-measurement states to be mixed ones. Could one then find a rule for choosing post-measurement states such that the second law would be satisfied in the above protocol?

While we are not in position to make a final claim at this moment, we nevertheless believe that these ideas reveal an intriguing aspect of generalized systems such as the hypercube bit. We hope that the above discussion will stimulate further work on thermodynamics in GPTs, which we feel may have impact on our understanding of quantum theory.


\emph{Acknowledgements.} N.B. acknowledges financial support from the Swiss National Science Foundation (grant PP00P2\_138917), SEFRI (COST action MP1006), and the EU DIQIP. M.K. acknowledges financial support from ANR retour des post-doctorants NLQCC (ANR-12-PDOC-0022-01). PS acknowledges support by the Marie Curie COFUND action through the ICFOnest program.

\end{document}